\documentstyle[12pt]{article}
\begin{document}
\def \be  {\begin{equation}}
\def \ee  {\end{equation}}
\def \beq  {\begin{equation}}
\def \eeq  {\end{equation}}
\def \ba  {\begin{eqnarray}}
\def \ea  {\end{eqnarray}}
\def \baa {\begin{eqnarray*}}
\def \eaa {\end{eqnarray*}}
\def \lab #1 {\label{#1}}
\newcommand\bqa {\begin{eqnarray}}
\newcommand\eqa {\end{eqnarray}}
\newcommand\pr {\partial}
\newcommand\apr {\overline {\partial }}
\newcommand\nn {\nonumber}
\newcommand \noi {\noindent}
\newcommand{\bear}{\begin{array}}
\newcommand{\enar}{\end{array}}
\newcommand{\hf}{\frac{1}{2}}
\newcommand{\vx}{\vec{x}}
\newcommand{\R}{\mathbb{R}}
\newcommand{\C}{\mathbb{C}}
\newcommand{\Q}{\mathbb{Q}}
\newcommand{\F}{\mathbb{F}}
\newcommand{\A}{\overline{\mathbb{C}}}
\newcommand{\Z}{\mathbb{Z}}
\newcommand{\bg}{{\bf g}}
\newcommand{\Tpl}{{T}_+}
\newcommand{\Tmin}{\mathcal{T}_-}
\newcommand{\LL}{{L}}
\newcommand{\AAA}{\overline{A}}
\newcommand{\inv}[1]{{#1}^{-1}} 

\def\t{\theta}
\def\T{\Theta}
\def\w{\omega}
\def\ov{\overline}
\def\a{\alpha}
\def\b{\beta}
\def\g{\gamma}
\def\s{\sigma}
\def\l{\lambda}
\def\wt{\widetilde}

\def \CO {{\cal O}}
\def \CP {{\cal P}}
\def \CT {{\cal T}}
\def \CM {{\cal M}}
\def \CK {{\cal K}}
\def \CH {{\cal H}}
\def \CI {{\cal I}}
\def \CV {{\cal V}}
\def \CJ {{\cal J}}
\def \CL {{\cal L}}

\font\cmss=cmss12 \font\cmsss=cmss10 at 11pt
\def\inbar{\,\vrule height1.5ex width.4pt depth0pt}
\def\IC{\relax\hbox{$\inbar\kern-.3em{\rm C}$}}
\def\IZ{\relax{\hbox{\cmss Z\kern-.4em Z}}}
\def\IR{{\hbox{{\rm I}\kern-.2em\hbox{\rm R}}}}
\def\R{{\tiny \IR}}
\def\IP{{\hbox{{\rm I}\kern-.2em\hbox{\rm P}}}}
\def\II{\hbox{{1}\kern-.25em\hbox{l}}}

\begin{titlepage}

\hfill\parbox{40mm}
{\begin{flushleft}  ITEP-TH-19/08\\
\end{flushleft}}

\vspace{10mm}

\centerline{\large \bf Amplitudes in  the N=4 SYM from  Quantum Geometry}
\centerline{\large \bf of the Momentum Space}

\vspace{17mm}

\centerline{\bf A. Gorsky}
\vspace{10mm}

\centerline{\it Institute of Theoretical and Experimental Physics}
\centerline{\it B. Cheremushkinskaya 25, Moscow, 117259, Russia}

\vspace{1cm}

\centerline{\bf Abstract}
We  discuss  multiloop MHV
amplitudes in the $N=4$ SYM theory in terms of
effective gravity in the momentum space with  IR regulator
branes as degrees of freedom.
Kinematical invariants  of  external particles
yield the  moduli spaces
of complex or Kahler structures which
are the playgrounds for the Kodaira-Spencer(KS)
or Kahler type gravity.
We suggest  fermionic
representation of the loop MHV amplitudes in the $N=4$ SYM theory assuming the
identification of  the IR regulator  branes with KS fermions in the B model
and Lagrangian branes in A model.
The two-easy mass box diagram is related to the  correlator
of fermionic currents
on the spectral curve in B model or hyperbolic volume
in the A model and it plays the role of
a building block in the whole picture. The
BDS-like anzatz has the interpretation as the semiclassical limit
of a fermionic correlator.
It is argued that   fermionic representation implies a kind of
integrability on the moduli spaces.
We conjecture the interpretation of the reggeon degrees of freedom in terms of
the open strings stretched between the IR regulator branes.



\end{titlepage}

\vspace{1mm}

\section{Introduction}

The $N=4$ SYM theory provides a possibility to recognize  some
features of the theories with less amount of SUSY. While
$N=4$ SYM  is far from the QCD-like theories in the infrared because of
the lack of confinement it shares  common features in  UV region
where  physics in asymptotically free theories is described
within a perturbation theory. That is considering the perturbative
expansion in $N=4$ SYM  coupling constant which does not run we
could try to clarify some generic properties  of the perturbative
expansion in the gauge theories.

It is of  prime importance  to discover any hidden symmetries at
high energies or equivalently hidden integrable structures providing
the nontrivial conservation laws which restrict the form of the
scattering amplitudes. In the four-dimensional setup  the
integrability behind the amplitudes is known  only at the Regge
limit when the $SL(2,C)$ spin chain gets materialized
\cite{lipatov1,fadkor}(see \cite{bbgk} for review).

The simplest objects  at  generic kinematics are the MHV
amplitudes which are the perfect starting point for any discussion
since at the planar limit they can be described in terms of the
single kinematical function. Even at the tree level MHV amplitudes
\cite{parke} enjoy some remarkable properties. They are localized on
the complex curves in the twistor space \cite{witten} (see \cite{cs}
for review) and can be described as the correlators of chiral
bosons on the genus zero Riemann surface \cite{nair}. It turns out
that the generating function for the tree MHV amplitudes is just the
particular solution to the self-duality equation in YM theory \cite{bardeen,rs}.
It substitutes the naive superposition of the plane waves of the
same chirality in a nonlinear theory.  Moreover this solution
provides the symplectic transformation \cite{gr}(see also
\cite{mansfield}) of the YM theory in the light-cone gauge
into the so-called tree MHV Lagrangian formulated in
\cite{mhv} which to some extend is the analogue of the t'Hooft
effective vertex generated by instantons. However this approach
becomes less clear when going to higher loops. Indeed, the attempt
to formulate the one-loop MHV amplitudes in a twistor-like manner  was not
successful enough \cite{csw} and certainly calls for additional insights on the problem.

One more line of development based on a first quantized picture for the loop
calculations was initiated in \cite{gopa1}. It was shown that the
three-point amplitude written in the Schwinger parametrization
implies the identification of the Schwinger proper time parameter
with the radial coordinate in the AdS geometry providing some
rationale for the appearance of the AdS space. The similar interpretation holds
true for the calculation of the one-loop effective actions in the
different backgrounds \cite{gorly}. However starting from one loop four-point
amplitude  the situation becomes more subtle because of the emerging
moduli space. It was suggested in   \cite{gopa3} that the Feynman
diagram can be presented in terms of the skeleton graph parameterized
by the set of Schwinger parameters. This set of  parameters can be
mapped for a planar limit of  n-point amplitude into the manifold $M_{0,n}\times R_{+}^n$
where $M_{0,n}$ is the moduli space of the n-punctured sphere. The mapping
of Schwinger parameters into the coordinates on the moduli space is
quite nontrivial and for instance does not respect the special conformal transformations
\cite{ahar1}. The gluing of the segments of the skeleton diagram
is subtle but some arguments based on OPE supporting this picture were presented
\cite{ahar2}. Hence we could expect that  loop amplitudes
in the SYM theory can be expressed
as correlators of some vertex operators on the moduli space
of the complex structures  which is the framework of the B model.

Another important starting point for the multiloop generalizations
was provided by the geometrical picture found in \cite{dav}.
It was argued that the one-loop amplitudes can be identified with
the hyperbolic volume of the ideal tetrahedron in the space
of the Feynman parameters. The corresponding Kahler moduli
are  fixed by the kinematical invariants. The hyperbolic volumes
of three-dimensional manifolds are the natural playground for the
A-model and the $SL(2,C)$ gravity provides the natural generalizations
of the one-loop answer. It is possible to consider the partition
function of $SL(2,C)$ gravity in terms of the proper  gluing the three-dimensional
manifold from the ideal tetrahedrons \cite{hikami}.

More recently Bern, Dixon and Smirnov (BDS) have formulated the  conjecture
\cite{bds} that all-loop MHV amplitudes get exponentiated and
factorize into  IR divergent and  finite parts. Moreover it
was conjectured that the finite part of all-loop amplitude involves
only all-loop cusp anomalous dimension $\Gamma_{cusp}(\alpha)$ and finite part of one-loop amplitude.
Inspired by this conjecture Alday and Maldacena have calculated the
amplitude at  strong coupling regime via  minimal surfaces in
$AdS$-type geometry with the proper boundary conditions \cite{am}.
They have found  unexpected relation between the MHV amplitudes in planar
limit of $N=4$ SYM theory and Wilson polygons in the momentum space.

The Wilson polygon-amplitude duality  refreshes the
problem but deserves for the explanation itself. It was originally
formulated at  strong coupling when the Wilson loop is calculated
in terms of minimal surface in the $AdS_5$ geometry upon a kind of
T-duality transform.  Later it was
shown that duality holds true at the perturbative regime as well
\cite{drum1,brand} which puts it on  more firm ground. Recently
the explicit derivation of the one-loop duality has been presented \cite{gz}.
The important
anomalous Ward identity for the special conformal transformations
with respect to the dual conformal group has been derived.
It fixes the kinematical dependence of the amplitudes up to five
external legs \cite{drum07,drum08}. However  Ward identities tell
nothing about the functional form of the amplitudes  starting
from six external legs. Recently the dual superconformal
group was identified as the symmetry of the worldsheet theory
of the superstring in $AdS_5\times S^5$ geometry \cite{bermal, beisert}.

Finally it was recognized that   BDS anzatz fails  at weak coupling
at two loop level for six external legs \cite{six1,six2} and at strong
coupling \cite{am2,mironov} for infinitely large number of external legs.
Moreover the BDS anzatz seems not fit well with the Regge limit
\cite{regge}(see however \cite{brower}). On the other hand at two loop level the duality between
Wilson polygon and MHV amplitude survives. The current status
of the whole problem has been  reviewed in \cite{alday}.

There  are a lot of pressing questions to be answered. Just  mention
a few;
\begin{itemize}

\item{Is there some geometrical picture behind the BDS-like anzatz which would
suggest the way of its necessary generalization?}
\item{Is there the generalization of the dual conformal Ward identity
which would fix the functional form of the one-loop amplitude for
any number of external legs?}
\item{Is there the fermionic representation for the loop amplitudes
which would imply the hidden integrability?}
\item{Is there clear geometrical picture behind the
reggeization of the gluon?}

\end{itemize}
In what following  we shall suggest the  answers to some of these
questions and  make a couple of conjectures.

To some extend we shall try to generalize the geometrical  picture for the tree
amplitudes suggested in \cite{witten}. At the tree level in \cite{witten}
the Euclidean D1 "instanton" branes with the attached open strings have been
considered in the twistor space. The D1 brane is localized
at the point in the Minkowski space in agreement with the
locality of the vertex generating tree MHV amplitude in the MHV formalism. To describe
the loop amplitudes we shall adopt a little bit different picture
and consider $C^4$ manifold in the B model. The  B branes
substitute "D1 instantons"  and are embedded in $C^4$.
The somewhat similar objects
were also introduced as the IR regulator branes in the
Alday-Maldacena calculation. Indeed the dilaton field
gets changed upon the T-duality in the RG radial coordinate which means
that D-instanton is added to the background. After the Fourier transform
along flat four-dimensions D-instanton gets transformed into the D3 brane
we shall work with.
The Wilson polygon which corresponds
to the boundary of the string worldsheet and is presumably dual to the amplitude is
located just on these IR regulator branes.
Contrary to the  previous considerations the positions
of the regulator branes will not be free but determined dynamically in terms of the
cross-ratios of the external momenta.

The emerging moduli space of IR regulator  branes
plays the central role in the picture. Contrary to the tree level the KS
degrees of freedom in the B model  defined on the moduli space do not decouple
and provide the phase space for
the corresponding integrable system.
The essential point in our approach concerns the quantization of the
emerging moduli spaces  and the
identification of the corresponding Planck constant
with some function of the YM coupling constant.
Therefore to some extend we could tell that the loop MHV amplitudes
emerge upon a kind of gravitational dressing of tree ones within the
Kodaira-Spencer type gravity in the "momentum" or twistor space.

The physics of the scattering at the loop level can be treated from the different perspectives.
From the point of view of the KS gravity on the moduli space we are calculating the correlator
of the fermions or the
fermionic currents  which can be identified with the
tau-function of the 2d integrable system. The second viewpoint
concerns a consideration of the  gauge theory on the IR regulator
branes whose number  is fixed by the number of the external
particles. Finally one could consider the worldsheet viewpoint
where the regulator branes provide the proper boundary conditions
for the string.
These viewpoints are complimentary and allow to check the self-consistency
of  approach.

Within the KS perspective we shall discuss the
fermionic representation behind the loop MHV amplitudes which
generalizes the Nair's fermionic representation for the tree
amplitudes. The fermionic picture is a heart of the integrability
which admits the representation in terms of the chiral fermions on
the Riemann surface in the external gauge field. The gauge field
on the Riemann surface represents the "point of Grassmanian" or in physical terms the
particular Bogolyubov transformation between the fermionic vacua.
Such emerging fermionic picture is known to be quite generic in
the set of examples which involves minimal string models
\cite{shih}, $c=1$ string \cite{seiberg} and crystal melting problem
\cite{crystal}. This approach was summarized in \cite{vafa} where
it was argued that fermions in the KS gravity correspond to mirror
of Lagrangian branes  in the A model. These B branes are also
refereed to as Kontsevich or
noncompact branes and their positions on the Riemann surface yield
the  "times" in the corresponding integrable systems. Note  that in
the framework of the topological strings in A-model we discuss the
Kahler geometry while in B-model the complex geometry is captured by
the Kodaira-Spencer \cite{kodaira} theory.

The natural question concerns the role of the Riemann surface
in the B model where the KS fermions lives on.
In the several examples  it collects the information about the
the infinite set of the anomalous Ward identities in the theory
\cite{vafa} and
encodes the unbroken part of the $W_{\infty}$ symmetry.
More qualitatively it means that if we introduce the  IR
regularization of the theory corresponding to the infinite blow-up of $C^3$
in the  A model the "anomaly" survives upon the
regularization  removed.
Note some analogy with the description of the Seiberg-Witten
solution to the low-energy $N=2$ SYM theory \cite{sw}. In that case
we have first to perform the summation over the point-like instantons \cite{nekrasov}
which amounts to the particular blow-ups  and desingularizes the target space theory geometry providing the
nontrivial Riemann surface. The physical correlators after all
are calculated in terms of the fermions on this emerging Riemann surface.

The fermion
one-point function corresponds to the Baker-Akhiezer
function in the integrability framework and to the single regulator
brane insertion at some point on the moduli space. Since generically we are interested in the quantum integrable system
the Riemann surface gets quantized and yields the corresponding Baxter equation
\cite{sklyanin}. The
semiclassical solutions to the Baxter equation  which are the generating functions
for the Lagrangian sub-manifolds in the particular integrable system  play
important role
in the analysis. They serve as the building blocks for the
correlators in the $N=4$ YM theory and can be considered as
the "semiclassical B brane wave
function" or  as the
effective action in the  gauge theory on the brane worldvolume.
From the moduli space  viewpoint the solution to
the Baxter equation provides the generating function of the
Lagrangian sub-manifold.
The natural integrable system on the moduli space can be identified with the  3-KP system
however similar to the $N=2$ SYM one could expect the
pair of integrable systems - 2D field theory and finite dimensional one.
The natural finite dimensional integrable system which is responsible for the hidden
symmetries at the generic kinematics is conjectured to be related
to the Faddeev-Volkov model \cite{volkov} and the corresponding statistical
model \cite{bms} based on the discrete quantum conformal
transformations.

Since  we are trying to sum  the perturbation series
the YM coupling constant is expected to be involved into
some algebraic structure behind  all-loop answers. It is this
hidden symmetry which provides the choice of the particular solution
to the Yang-Baxter equation. The Faddeev-Volkov solution to
the Yang-Baxter implies that we are actually trying to relate the
YM coupling constant with the parameter $q$ of $U_{q}(SL(2,R))$.
The proper identification turns out to be nontrivial problem since
in particular
it has to respect the S-duality group in $N=4$ theory. It will
be argued that the BDS anzatz corresponds to the limit
$q\rightarrow 1$ while the Regge limit seems to be related
to the opposite "strong coupling regime" of the quantum group.

The consideration of the gauge
theories on the regulator brane worldvolume is useful as well. The
theory can be thought of as  in the Coulomb phase and the position of the regulator brane in
the particular complex plane corresponds to the
coordinate on the Coulomb moduli space.
Since all regulator  branes are at different positions on the moduli space the
theory generically has the gauge group $U(1)^k$ where k is related
to the number of the external gluons however there are possible
enhancements to the nonabelian factors at some kinematical regions.
The effective action of each $U(1)$ gauge theory on the regulator brane
plays the role
of the wave function of the two-dimensional fermion
or B brane  in KS gravity on the B model side.
Similarly the worldvolume  theory on the Lagrangian regulator D2
branes can be considered on the A-model side. In this case we shall
consider the twisted superpotentials in the worldvolume theory.
The minimization of the effective superpotential amounts to the
selection of the positions of the Lagrangian branes at the base manifold.

The Riemann surface involved has the interpretation in the regulator brane
worldvolume theory as well.
To this aim note that the coordinate on the moduli space
plays the role of the complex scalar in the B brane worldvolume theory.
One can consider the change of variables in the theory
corresponding to the reparametrization of this scalar field.
Such transformation is familiar in the $N=1$ SYM theory as the generalized
Konishi transformations which are anomalous. One can collect all Konishi
transformations or Virasoro constraints in the Dijkgraaf-Vafa
approach \cite{dv} into the single generating equation which yields
the particular Riemann surface   \cite{chiral,gorsky2}.
It can be considered as  analogue of the chiral ring
in the worldvolume theory on the regulator branes.

It is important to discuss separately the special Regge kinematical
region where the hidden symmetries of the amplitudes were found
for the first time. The hidden symmetries were captured at one loop
by the $SL(2,C)$ spin chains \cite{lipatov1,fadkor}. It was shown in \cite{gkk}
that the N-reggeon dynamics belongs to the same universality class
as conformal $N=2$ SQCD with $N_f=2N$ at the strong coupling orbifold point.
We shall argue that the brane geometry in the  reggeon case is similar
to the one in SQCD which provides  the qualitative
explanation of the same universality class for both theories.
The new object  is the open
string stretched between two regulator branes and is the
analogue of the massive vector bosons and monopoles in the conventional $N=2$ SYM
theory. Here we shall tempt to interpret these open strings
as the "reggeons". The "masses" of these effective degrees of freedom
correspond to the differences of the positions of the
regulator branes on the proper Riemann surface and therefore
depend on the kinematical invariants. We shall comment
on  the possible
link of this picture with the effective Reggeon field theory
\cite{lipatoveff}.

The paper is organized as follows. In Section 2 we remind the main
features concerning the loop MHV amplitudes. In Section 3 we briefly
consider the example of the  $c=1$ noncritical string which
provides some intuition on the calculation of the amplitudes
from the target space perspective. In Section 4  we review
the relevant properties of the quantum dilogarithm. Section 5 is
devoted to the formulation of
our conjecture for the $N=4$ MHV amplitudes.  In Section 6 we make
conjecture on the pair of the underlying integrable systems. Some arguments
concerning   the  Regge limit of the amplitudes are present in Section 7.
In the last Section  we collect
the main points of our proposal and fix the open questions to be
answered.

\section{The  loop results for the MHV amplitudes}

Let us remind the main results concerning the loop MHV amplitudes.
The  MHV gluon
amplitudes  involve two gluons of the negative chiralities and
the rest of gluons have  positive chiralities. Consider the ratio of
all-loop and tree answers. The following form of the all-loop amplitudes has been
suggested in \cite{bds}

\beq log(\frac{M_{all=loop}}{M_{tree}})= (F_{div}+
\Gamma_{cusp}(\lambda) M_{one-loop}) \label{af}
\eeq
which involves
only the all-loop answer for the cusp anomaly $\Gamma_{cusp}$ and
one-loop MHV amplitude. The IR divergent part $F_{div}$ gets
factorized in the all-loop answer. The cusp anomaly measures  UV
behavior of the contour with cusp \cite{polyakov}. Recently the
closed integral equation has been found for the cusp anomalous
dimension in $N=4$ SYM theory \cite{bes} which correctly reproduces
the weak and strong coupling expansions.

The finite part of the one-loop MHV which presumably defines the all-loop
answer  can be written in terms of the finite part of the
so-called two-mass easy box function $F^{2em}$ \cite{brand}

\beq M_{one-loop,finite}= \sum_{p,q} F^{2em,f}(p,q,P,Q)\eeq This
function can be expressed in terms of the dilogarithms only

\beq F^{2em,f}(p,q,P,Q)= Li_2(1-aP^2) + Li_2(1-aQ^2)
- Li_2(1-a(q+P)^2) - Li_2(1-a (p+P)^2) \eeq
where

\beq a=\frac{P^2 + Q^2 -(q+P)^2 -(p+P)^2}{ P^2 Q^2 -(q+P)^2(q+P)^2}
\eeq
and $p+q+P+Q=0$. One more expression for the function $F^{2em,f}$
can be written in terms of the variables
$x_{i,k}=p_i-p_k$ as  the sums \cite{drum1}

\beq \sum_i \sum_r Li_2 ( 1-
\frac{x_{i,i+r}^2x_{i=1,i+r+1}^2}{x_{i,i+r+1}^2x_{i-1,i+r}^2})\eeq
where
\beq
x_i=p_{i+1}-p_i \eeq
Since all external momenta are on the mass shell the arguments of
dilogarithms are expressed in terms of the cross-ratios of the
scalar products of the momenta only.

Since we shall aim to get the geometrical interpretation of the BDS
anzatz we need first the clear geometry behind one loop. It is provided
by the observation in \cite{dav} that box diagram with all external
particles off-shell just calculate the volume of the three-dimensional
ideal hyperbolic tetrahedron in the space of the   Feynman parameters
(see Appendix).
The Kahler modulus $z$ of the tetrahedron  is fixed by the kinematical
invariants and in the two-mass easy box it reduces to the conformal ratios.
The appearance of the hyperbolic volume implies that the topological string approach
or CS with $SL(2,C)$ group are relevant \cite{hikami}. Indeed we can
consider the ideal tetrahedron
as the knot complement and  calculate its volume via the Chern-Simons theory action
with the complex group. To some extend the exponentiation of the one-loop
answer in the BDS manner corresponds to the calculation of the
classical partition function in $SL(2,C)$ CS action which involves $exp(Vol(z))$ factor.
Let us emphasize that  the volume is finite if we consider off-shell particles
particles only and IR divergence of the amplitude corresponds to the
divergence of the volume. Some development along such interpretation
of the higher loops can be found in \cite{brod}.

The topological string picture usually can be represented both in the A-model
and B-model sides. The hyperbolic volume calculation evidently corresponds
to the Kahler gravity in A-side so one could ask about the one-loop geometry
in the B-model language. It can be uncovered indeed if we recall that dilogarithm
is the natural object on the moduli space $M_{(0,n)}$ \cite{fock99}
where it provides the proper canonical transformations.
The natural arguments of dilogs as functions on the moduli space
are just the conformal ratios
appeared in the one-loop integrals.
We shall try
to interpret  BDS-like structure in the B-model language as the result of the quantization
of the Teichmuller space when the one-loop dilogs are substituted by the
quantum dilogs. This viewpoint will be useful when we shall search for the
proper "degrees of freedom" involved into higher loops calculations.
They will be conjectured to be IR regulator branes or equivalent fermions.

\section{The c=1 Example}

The useful example which shares some essential features with our
problem is provided by the $c=1$ model. The noncritical  $c=1$
model describes the
string in the one dimensional compact target space and because of
the non-criticality the target geometry becomes
two-dimensional due to the Liouville direction. The only physical
modes are  massless tachyons generically gravitationally dressed
by the Liouville modes.

The theory enjoys two
natural types of branes which provide the boundary conditions
for the strings; so
called ZZ and FZZT branes. The  compact ZZ branes correspond
to the unstable D0 branes localized in the Liouville direction. On
the other hand the  noncompact branes correspond to
the stable D1 branes extended along the Liouville coordinate till
the Liouville wall  at the cosmological constant scale $\mu$
\cite{seiberg}.

The explicit target space description is more appropriate for  our
purposes. This approach is natural in the topological string setup
and was developed in \cite{seiberg}. The crucial point is the
existence of the so-called chiral ring in the theory. In the c=1
case it collects the information
about the set of certain anomalous relations in the theory. The most
useful description of the chiral ring  involves the Riemann surface
supplied with some meromorphic differential. For $c=1$ theory it reads
as
\beq x^2 -y^2=const \eeq
where $x,y\in C$.
In terms of the
Riemann surface a compact ZZ brane corresponds to the tunneling state
in the inverted oscillator or equivalently to the closed cycle on
the surface pinched at the degeneration point. On the other hand
noncompact  FZZT brane corresponds to the open path on the surface.
The corresponding semiclassical ``wave function''
\beq
\Psi_{FZZT}\propto exp(i\int ydx)
\eeq
transforms nontrivially
when going to different coordinate paths on the surface.

Another useful language is provided by the  matrix model approach.
The random matrix model is usually built on the set of infinite
number of ZZ branes whose coordinates correspond to the eigenvalues
of the matrix of the infinite size triangulating
the string worldsheet. On the other hand one can
consider the matrix model of the Kontsevich type on the N noncompact
FZZT branes. In this case one deals with the $N\times N$ matrix
model with the source term $Tr\Lambda X$ and the eigenvalues of the
matrix $\Lambda$ encode the positions of the FZZT branes. The
appropriate object in the matrix model is its resolvent
\beq W(z)=<
Tr \frac{1}{z-M}> \eeq
obeying the loop equation which in the
semiclassical limit coincides with the equation of the spectral
curve.

All languages yield  the same important feature of c=1 model - its
hidden integrability. It turns out that tau-function of the Toda
hierarchy serves as the generating function for the tachyonic
amplitudes \cite{tach}. This objects can be naturally described in terms of the
chiral boson $\phi(z)$ on the spectral curve or as the corresponding
fermion \beq \Psi(z)=exp(g_s^{-1}\phi(z))\eeq

More precisely one considers the following matrix element in the
theory of the chiral boson or its fermionized version
\beq
\tau(t,A)=<t|exp(\psi A \psi)|0> = <t|V>
\eeq
where
$|t>=exp(\sum_k t_k \alpha_k)$ is generically the coherent
state of the chiral boson
with modes $\alpha_k$ on our Riemann surface. The matrix A encodes
the scattering of fermions off the Liouville wall which essentially
provides the whole answer for the tachyonic amplitudes \cite{tach}. The
integrability encodes the infinite number of the Ward identities in
the theory followed from the symplectic invariance of the Riemann
surface which is the complex Liouville torus for the complex
Hamiltonian system. Some part of the symmetries is spontaneously
broken yielding the corresponding Ward identities. Some of them are
unbroken yielding the equation of the quantum spectral curve
\cite{vafa}. The set of Ward identities can be formulated in terms
of the fermionic bilinears on the surface and provides the exact
answers for the correlators in the theory.

The FZZT or noncompact branes parameterize the moduli space of the
complex structures in the target geometry and can be naturally
treated within the KS gravity in the target space.
The positions of the noncompact FZZT
branes $z_i$ yield the following "times" in the Toda integrable system

\beq T_k=\frac{1}{k}\sum_{i=1}^N z_i^{-k}\eeq

The "wave function" of the FZZT brane itself can be considered as the Baker-Akhiezer
function in the integrable systems. Quantum mechanically
the Riemann surface gets quantized yielding the
Baxter equation for the eigenvalue of the Baxter operator.
Solution to the Baxter equation corresponds to the wave function of the single
separated variable. The brane interpretation of the separated
variables has been suggested in \cite{gnr}.  The correlator of the Kontsevich
branes inserted at points $z_i$ has the following structure

\beq
<0|\Psi(z_i)\dots \Psi(z_n)|V>=\prod_{i,j}(z_i
-z_j)e^{\sum _{i}\phi(z_i)}
\eeq
where one can clearly  distinguish the
classical and quantum components of the answer.
Another point to be mentioned is the identification of the
quantization parameter. In the commutation relation on the quantized
Riemann surface
\beq [x,y]=g_s \eeq
the Planck constant is just the
string constant or the graviphoton field.

To summarize, $c=1$ model provides the example when the
non-perturbative structure of
the theory is stored is the Riemann surface in the B model which
collects the information about the chiral ring. This Riemann surface
has to be considered as the energy level of some complex Hamiltonian
and the set of canonical transformations of the phase space
amounts to the set of Ward identities in the initial model which
fix the scattering amplitudes. The deformations of the complex
structures which are dynamical degrees of freedom in the
Kodaira-Spencer theory are parameterized by the fermions on the
Riemann surface which upon quantization yield the Baxter equation
of the corresponding integrable system.

\section{Quantum dilogarithm}
The one-loop answer is expressed in terms of the dilogaritms
hence  in this Section we shall briefly review some relevant
properties of the quantum dilogarithm
defined as the following integral

\beq\Psi_{b}(z) = exp(\frac{1}{4}\int \frac{e^{-2izx}dx}{x sinh(bx)sinh(b^{-1} x)})
\eeq
The integration contour is chosen in such way that the integral reduced to the infinite sum via
residue calculation
\beq \Psi_{b}(z)= exp(\sum_{n}\frac{e^{-nx}}{n[n]}) \eeq
It  obeys the functional equations
\beq
\Psi_{b}(z)\Psi_{b}(-z)= e^{i\pi z^2 -i\pi (1+2c_q^2)/6}
\eeq
and
\beq
\Psi_{b}(z-b^{\pm 1}/2)= (1+ e^{2\pi zb^{\pm 1}}) \Psi_{b}(z+b^{\pm 1}/2)
\eeq
where $c_b=\frac{i}{2}(b+b^{-1})$,
as well as the unitarity condition
\beq
\bar{\Psi}_b(z)= \Psi_b(\bar{z})^{-1}
\eeq

The quantum dilogarithm can be represented as the ratio
of two q-exponentials

\beq
\Psi_b (z)=\frac{(e^{2\pi (z +c_b)b};q^2)_{\infty}}
{(e^{2\pi (z -c_b)b^{-1}};\tilde{q}^2)_{\infty}}
\eeq
where $q=e^{i\pi b^2},\quad \tilde{q}=e^{-i\pi b^{-2}}$ and
\beq
(x,q)_{\infty}=\prod_{n}(1-q^n x).
\eeq
The dilogarithm enjoys the duality

\beq
\Psi_{b} (z)= \Psi_{b^{-1}}(z)
\eeq
which is essential when it is involved  into gluing of
the conformal blocks in the Liouville model \cite{teschner,kashaev2}.
Note that the central charge in the corresponding Liouville theory
reads as
\beq
c_{liouv}=1 +6(b+b^{-1})^2
\eeq

The quantum dilogarithm is natural object from the viewpoint
of the quantum torus algebra defined by the relation

\beq \hat{U}\hat{V}=q\hat{V}\hat{U}
\eeq
where $\hat{U}=exp(i\hat{x})$ and $\hat{V}=exp(i\hat{p})$. It is
assumed that the  variables $x,p$  obey the canonical phase space commutation relations. In terms of
the quantum torus the quantum dilogarithm is defined via the
relation
\beq \Psi(\hat{V})\Psi(\hat{U})=
\Psi(\hat{U})\Psi(-\hat{U}\hat{V})\Psi(\hat{V})
\eeq
This  property represents the so-called quantum pentagon relation 
\cite{fk} and reduces to the classical Rogers
identity for $Li_2(z)$ in the semiclassical limit.

Quantum mechanically the dilogarithm defines the operator
with the kernel
\beq K(x,z)= \Psi_q(z)e^{\frac{-zx}{2\pi q}}\eeq
which serves as the generating function of the following canonical
Backlund type transformations

\beq U\rightarrow ((1+qU)V \qquad V\rightarrow U^{-1} \eeq  which
belongs to the outer automorphisms of the algebra of
functions on the quantum torus. The important property of this
transformation  is the operator version of the pentagon relation
\beq \hat{K}^5 =1 \eeq which states that automorphism is of the
fifth order.

Dilogarithm  plays an essential role in the symplectic treatment
of the Teichmuller space. Remind that the Teichmuller space $T(S)$ is
the space of the complex structures on the Riemann surface S
modulo trivial diffeomorphisms
homotopy equivalent to the identity while the moduli space $M(S)$ is obtained
via the factorization of the Teichmuller  by the action
of the mapping class group. The quantum dilogarithm plays
the  role of the quantum generating function for the particular
element of the
mapping class group - flip transformation. The flip transformations
are responsible for the  generic transition maps between
coordinate systems corresponding to the different triangulations.
The quantization of the Teichmuller space was developed in
\cite{kashaev,fock99}(see \cite{teschner} for review).

There are different ways to introduce the coordinates on the
Teichmuller
space of the punctured spheres we are interested in. Of particular interest are the
coordinates related to the geodesic lengths and the corresponding
classical Poisson structure can be written in a simple way in terms
of triangulations. These shear coordinates can be introduced in terms
of the cross-ratio of the four points on the real line connected by
the geodesic circles in the upper half-plane
\beq
e(z)=\frac {(x_2-x_1)(x_4-x_3)}{(x_3-x_2)(x_4-x_1)}
\eeq
The natural arguments of the dilogarithms defined
on the moduli space
are just cross-ratios which we meet in the
answers for the MHV amplitude.
In the semiclassical limit we have
\beq 
\Psi_{b} (x)\rightarrow exp(\frac{1}{b}Li_2(e^x))
\eeq

In the context of integrability dilogarithms appear as the ingredients
of the fundamental R-matrix involved into the description of the
Liouville and sin-Gordon theories in the discrete space-time. In terms
of the discretized version of the Kac-Moody currents with the commutation
relation similar to the quantum torus
\beq
\omega_n \omega_{n+1}=q^2 \omega_{n+1} \omega_{n}
\eeq
\beq
\omega_n \omega_{m}= \omega_{m} \omega_{n},\qquad
|n-m|\geq 2\eeq
The periodicity condition for the current variables is
assumed $\omega_n = \omega_{n+N}$. In terms of these dynamical
variables one can define the solution to the Yang-Baxter
equation depending on the spectral parameter $\lambda$ \cite{faddeev94}
\beq
R(\lambda,\omega))=\frac{\Psi_b(\omega)\Psi_b(\omega^{-1})}
{\Psi_b(\lambda \omega)\Psi_b(\lambda \omega^{-1})}
\eeq
The product of the R-matrixes over the lattice cites yields
the evolution operator for the massive model on the space-time lattice.

The finite-dimensional system discussed in the B-model framework
should carry some information on the S-duality of the $N=4$ gauge theory.
The modular parameter of the gauge theory
can be identified with the quantization parameter
of the integrable system. However to get the full modular symmetry
the  modular double \cite{faddeev99} has to be included into the game.
It unifies  two quantum tori with the
S-dual moduli. It turns out that the modular double plays the
crucial role in the quantum integrable systems providing
the self-consistency of the local and nonlocal integrals of
motion.

That is  integrable system has to be supplied with the
following symmetry $U_{q}(SL(2,R)\otimes U_{\tilde{q}}(SL(2,R)$.
The corresponding R-matrix acting  of the modular double
reads as
\beq R= e^{\frac{\pi}{2}(p_3 +p_2)\otimes ( p_1+p_4) }
\Psi(p_{13})\Psi(p_{34})\Psi(p_{23})\Psi(p_{24})\eeq
\beq
p_{ik} =p_i\otimes I +I\otimes p_k
\eeq
and the variables $p_i,i=1\dots 4$ obey the commutation relations
\beq
[p_k,p_{k+1}]=-2\pi i I
\eeq
The integrable system with such symmetry was found in \cite{volkov}.
It was argued in \cite{bms} that using the R matrix
for the modular double one can define the
positive weights which provide the unitary model.

Such type of R-matrix emerges
naturally within the discrete quantum Liouville
theory related to the discrete conformal transformations
\cite{volkov,bms,kashaev}. It was argued that
the structure of the modular double is necessary for the
self-consistent description of the Liouville theory
at the strong coupling region $1<c<25$. The integrability
of the model to some extend is equivalent to its very
quantum existence \cite{bms}. The link with the dilogarithms
involved into the description of the Teichmuller space goes as follows. The
universal Teichmuller space is known to be identified with
the coadjoint Virasoro orbit. On the other hand the Liouville action
plays the role of the geometrical action on this orbit \cite{shatashvili}
therefore it is no surprise that the triangulation
of the moduli space is related with the discrete quantum Liouville theory.

In the physical setup the quantum dilogarithm corresponds to the
probability of the  charged pair creation
in the constant external field  in four-dimensional scalar QED theory.
It is assumed that both electric (E) and magnetic (H) fields
are switched then the one-loop effective action reads as
\beq
L_{one-loop}=\frac{1}{16 \pi^2} \int \frac{dt}{t} e^{-m^2 t}(\frac{e^2 ac}{sinh(ect)sinh(eat)} - \frac{1}{t^2} - \frac{e^2}{6}(a^2 -c^2)) \eeq
where $a^2-c^2= E^2 -H^2$ and $ac=EH$. The last two terms provide
the proper substraction to get the finite answer.
The "strong coupling limit" $|b|\rightarrow 1$  of the quantum dilogarithm
corresponds to the almost self-dual external field  while the semiclassical
limit corresponds to the strong deviation from the self-dual regime.

\section{Finite part of $N=4$ SYM MHV amplitudes and "momentum space" geometry}

\subsection{The brane picture}
Let us now  formulate our
proposal for finite part of the MHV loop amplitudes.
Recall that the tree amplitudes were described in terms of the
D1  string instanton embedded into the twistor manifold \cite{witten}.
The instanton is localized at point in the Minkowski space and open strings
representing gluons are attached to  it. To describe the loop
amplitude we shall substitute D1 brane by the IR regulator  brane
embedded into the proper manifold. The gluons are attached
to the regulator branes whose embedding coordinates
are considered as dynamical degrees of freedom.
Contrary to tree case  regulator branes are localized at the sub-manifold of the
complexified Minkowski space. The loop amplitudes can be considered from the different
perspectives: in terms of the KS gravity on the particular Riemann surface,
within the worldvolume theory on the regulator
branes and in the theory on the string worldsheet.
Let us emphasize that the embedding of the IR regulator branes
nontrivially depends on the external momenta.

The starting
point is the representation of the $N=4$ theory via geometrical
engineering \cite{vafageom} as the IIA superstring compactified on the
three-dimensional Calabi-Yau manifold which was identified as  the
$K3\times T^2$ geometry in the singular limit. One has to
consider the singular limit of K3 manifold when it develops $A_{N-1}$
singularity, where N becomes the rank of the gauge group, and upon
blowing up procedure it can be represented as $ALE_N$ geometry. On
the other hand the Kahler class of the $T^2$ can be identified with
the coupling constant
\beq Area(T^2)=1/g_{YM}^2 \eeq
At weak
coupling  the torus is large and can be approximated by the complex
plane.  That is the geometry can be roughly approximated by $C^3$
upon the particular blow-ups.

As we have seen the one-loop answer for the MHV amplitude
determining  the BDS form of the amplitude involves the sum
of the dilogarithms depending on the cross-ratios of the
$x_i$ variables. Below we shall try to explain how such
functions  with cross-ratio arguments emerge naturally both in A-model
and B-model frameworks. As is well-known the A-model captures
the information about the Kahler moduli while the B-model
about the complex moduli and we shall see where these moduli comes from.
The brane description of the scattering amplitude involves
the set of the Lagrangian branes in the A-model and the
corresponding B-model branes. It is these branes which provide the
corresponding moduli spaces.

\subsubsection{B-model}
First we shall discuss B-model approach.
There are two natural B-model setups. The first one follows from
the topological S-duality \cite{sdual} and corresponds to the
same manifold with the S-dual moduli.
Since in our case area of the torus is defined in terms
of the YM coupling constant we end up with the  small
dual torus in the S -dual model at weak coupling. This topological B-model
in the S-dual geometry can be described in terms of the
noncommutative $U(1)$ gauge theory in D=6 which is naturally
defined on the D5 brane worldvolume.
Another viewpoint is provided by the mirror symmetry which maps
A-model to B-model on the different manifold. It is convenient to
consider the dual mirror geometry upon the infinite blowup of $C^3$.

Let us interpret the BDS anzatz in
terms of the correlator of the noncompact Euclidean B branes
embedded into the four dimensional complex space. Consider
3d complex manifold  which is mirror to the  topological
vertex \cite{vertex}. This manifold  classically is
described by the equation in the  $C^4$ with coordinates $x,y,u,v$

\beq xy=e^{u} +e^{v}+1 \eeq
At the discriminant locus it defines  the Riemann surface
\beq H(v,u)= e^{u}+e^{v}+1=0
\label{surface}
\eeq
of genus zero with three different asymptotic regions.

This Riemann surface emerges from the infinite
blow-ups of the origin of the toric fibration of $C^3$ upon the
mirror transform and provides the part of IR regularization of the
theory.
We shall try to argue that the loop MHV amplitudes can be identified with the
fermionic correlators on the Riemann surface (\ref{surface}). Fermions on the surface
(\ref{surface}) represent the degrees of freedom in the KS gravity.
They are identified with the IR regulator  branes imbedded into
$C^4$ geometry.

There are two B branes  defined by the equations

\beq
x=0 \quad H(v,u)=0 \eeq
and
\beq
y=0 \quad H(v,u)=0 \eeq
which  intersect along the Riemann surface.
The intersecting branes provides
the natural fermionic degrees of freedom on the intersection surface
\cite{dv07} from the
open strings stretched between them.
The fermions are in  external field amounted from the worldvolume
gauge connections on the intersecting branes. This gauge field
representing the point of the Grassmanian which can be read off
from the topological vertex. In addition to  two  branes
intersecting along the Riemann surface we introduce the set of Kontsevich -like
branes classically localized at the points $(v_i,u_i)$ at the Riemann surface.
The number of such branes is fixed by the number of the external gluons
and the coordinates of these branes on the surface are defined by some particular
cross-ratios.
The cross-ratios are the natural coordinates
on the moduli space of the punctured spheres that is the $(u,v)$ space
is related to the $T^{*}M_{0,4}$.
Hence we are in the framework
of the KS gravity and the fermions on the Riemann surface represent
the KS gravity degree of freedom.

The
Riemann surface gets quantized and the branes-fermions should obey
the equation of the quantum Riemann surface that is
Baxter equation which provides
the wave functions depending on the  separated variables \cite{gnr}.
The Baxter equation in our problem reads as
\beq (e^{\hbar \partial_v} +e^{v} +1)Q(v)=0 \eeq
Its solution corresponds to the vev of fermionic bilinear $J$
\beq Q(v)=<0|J(v)|V> \eeq
and turns out to
be the  quantum dilogarithm \cite{vafa}. Note that the solution to the Baxter
equation in our case  can not be presented in the polynomial form that is we have
infinite number of the Bethe roots.

To get the MHV all-loop amplitude in the BDS form we take the
semiclassical limit of the fermionic correlator on this surface.
Indeed using the semiclassical limit for the quantum dilogarithm
we can represent the four-point fermionic current correlator as

\beq <\bar{J}(z_1)\bar{J}(z_2) J(z_3)J(z_4)>\propto exp(\hbar^{-1}(
Li_2(z_3)+ Li_2(z_4)-Li_2(z_1)-Li_2(z_2))
\eeq
This expression exactly coincides   with the expression for the
finite contribution of the single 2-easy mass box diagram
hence upon  the identification of the Planck constant

\beq \hbar^{-1}=\Gamma_{cusp}(\lambda) \eeq
we reproduce BDS anzatz for the finite part of the amplitude. Indeed the one-loop
answer for the MHV amplitude can be expressed purely in terms of the sum
of 2-mass easy box diagrams with different grouping of the gluon momenta
and therefore in terms of the fermionic correlators.

Since the regulator brane ( D1 "instanton") yielding the tree amplitude is localized in
the complexified Minkowski space $M^c$ \cite{witten} one could ask about similar localization of regulator branes
responsible for the higher loop calculations. To this aim recall that
the complexified Minkowski space  $M^c$ is equivalent
to the Grassmanian $Gr(2,4)$. On the other hand the factor
of the Grassmanian by the maximal torus action is related to the
compactified moduli space \cite{kapranov}
\beq
Gr(2,4)// T =\bar{M}_{0,4}
\eeq
This representation allows to represent the complexified
Minkowski space
itself as the fancy divisor of the $M_{0,4}$ \cite{altmann}.
We suggest that this realization implies the localization
of the regulator branes on the submanifold of
$T^{*}(M^c//T)$ . It is natural to identify
this manifold with the Riemann surface where the KS degrees
of freedom live.

\subsubsection{A-model}
Let us present the qualitative arguments  concerning
the corresponding A-model picture.
In the A-model we introduce  the set of Lagrangian branes
with topology  $S^1\times R^2$. They can be also thought of as D6
branes if we add the conventional $R_{3,1}$ piece of geometry.
The emergence of the dilogarithm
as the wave function of the Lagrangian brane has been discovered
in the $C^3$ geometry in \cite{saulina}.
The brane/antibrane  can be considered as the insertion of the
fermion/antifermion \cite{saulina}  in the fermionic representation of the
topological vertex picture \cite{vertex}.

In this case we get the Kahler gravity as the target space
description of our geometry. The Lagrangian of the Kahler gravity
was conjectured to reduce to the $SL(2,C)$ CS theory on the
Lagrangian branes which describes the quantum geometry of the
hyperbolic space.  Since we are trying to identify the  amplitude
as the wave function of the Lagrangian branes, its argument
in the proper polarization should
be Kahler modulus of the ideal tetrahedron. This is indeed consistent
with the loop calculations
since the box diagram yields the hyperbolic volume in the space of the
Feynman parameters \cite{dav}.

The intersecting Lagrangian branes  naturally
provide the set of knots. The knot complements are the natural
hyperbolic manifolds and we can consider the triangulation of the
three- dimensional cusped hyperbolic space  by the ideal tetrahedron ( see, for
instance \cite{hikami}).
It turns out that quantum dilogarithm with the proper
argument can be attributed to each
tetrahedron and the main problem turns out to be the gluing
the whole  manifold from the several ideal tetrahedra. The gluing
conditions have the form of the Bethe Anzatz -type equations
if one attributes to the each tetrahedron a kind of
S-matrix \cite{hikami}. That is in the A-model picture the
all-loop answer deserves the accurate gluing of the submanifolds
in the hyperbolic spaces.

\subsection{The regulator  brane worldvolume theory}

Since fermions in KS framework are identified as the  regulator
B branes the natural question
concerns their four-dimensional worldvolume theory.
The theory on the regulator branes share many features
with $N=2$ and $N=1$ SYM low-energy sectors. The number of the regulator branes
is fixed by the number of the external gluons  so naively one could expect
a kind of $SU(K)$ gauge theory. The worldsheet theory on the regulator branes
enjoys the complex scalar
corresponding to  the complex coordinate $z$ of the brane on the Riemann surface (\ref{surface}).
This is similar to the situation when the vev of the scalar field corresponds
to the position of the D4 branes on the $u$-plane in the IIA realization
of the $N=2$ SYM theory \cite{wittenm}.

Since the different regulator branes are at the different points
on the Riemann surface  we can speak about the Coulomb branch
of the regulator worldvolume theory. However their positions on the
Riemann surface are fixed  that is we could say about the localization
of the B branes at the points of the moduli space $M_{(0,4)}$. Similar to the $N=1$
SYM theory when branes are localized at positions corresponding to
the discrete vacua  the  regulator branes are localized at some
points parameterized by the cross-ratios. These
points correspond to the local rapidities in the framework of
integrability and simultaneously have to
correspond to the minima of the effective superpotentials $W_{eff}(z_i)$
in the regulator worldvolume theory.

Since we identify dilogaritms as  the regulator brane wave functions  it is necessary
to explain where they come from in the worldvolume theory. The qualitative arguments
looks as follows. In the worldvolume theory there are massive excitations corresponding
to the open strings stretched between two regulator branes.  They are analogue of
the massive W-bosons in $N=1$ SYM theory on the Coulomb branch. In our case the masses
of these "particles" are related to the cross-ratios. To recover the dilogaritms let us remind
that usually in the external field the effective action develops the imaginary part
corresponding to the pair creation. The probability of the pair creation in the
external field is described by the classical trajectory in the Euclidean space
and in the leading approximation reads as
\beq
Im S_{eff}\propto e^{-\frac{m^2}{eE}}
\eeq
for a particle of the mass m in the external field E. Upon taking into account the multiple
wrapping and the quadratic fluctuations one gets for the scalar particle Schwinger pair
production
\beq
Im S_{eff}\propto \sum_{n} \frac{1}{n^2}e^{-\frac{nm^2}{eE}}
\eeq
that is dilog plays the role of the decay probability. Hence one can
say that we are considering the Euclidean version of the regulator
worldvolume theory and the amplitude from this viewpoint is described
via bounce type configuration corresponding to the creation of the
pairs of the effective massive degrees of freedom. Note that the real part
of the effective action corresponds to the summation over
the loop contributions
of the same degrees of freedom.

We have a $N=1$ type theory on the regulator branes with the
finite number of vacua
and the complex scalar field whose vacuum expectation
value corresponds to the coordinates of the B branes on the
Riemann surface.
Hence we can discuss the role of the symmetries implied
by the area preserving diffeomorphisms of the Riemann surface. From
the regulator worldvolume theory this transformation is the symmetry
of the target space analogous to the transformations of the
scalar field in $N=1$ gauge theory with adjoint scalar. It $N=1$ SYM case
the generalized Konishi anomaly
responsible for these transformations \cite{chiral,gorsky2}
\beq
\Phi\rightarrow f(\Phi)
\eeq
captures the unbroken part of $W_{\infty}$.

In the A-model one can similarly consider the worldvolume
theory on the D2(or D6) Lagrangian regulator branes. In this case the
corresponding dilogarithm functions emerge upon summation over
the disc instantons with boundaries located at the corresponding
Lagrangian branes
\beq
W_{eff}\propto \sum_{n} \frac{d_n}{n^2} e^{-nA}
\eeq
where A-is the corresponding area of the target disc.
The effective twisted superpotentials
in the three-dimensional  worldvolume theory with one compact dimension
were discussed in \cite{nekshat}. Roughly speaking the wave function of  D2 brane
has the form
\beq
\Psi(z)\propto e^{W_{twisted}(z)}
\eeq

Note that in
the A model   D2 branes can be considered as wrapped around the
ideal tetrahedrons whose Kahler modulus are defined by the cross-ratios
providing the masses of the same effective "W-bosons" as in B-model.
The issue of the gluing of the tetrahedra gets reformulated in terms
of the minimization of the total twisted superpotential which
is equivalent to the solution to the Bethe Anzatz equations
in the XXZ spin chain model \cite{nekshat}. The solution
to the Bethe Anzatz equations fixes the positions of the
Lagrangian branes. To fit this arguments with the complex moduli
remark that Bethe Anzatz equations for the XXZ model
is related to the solution to the classical equations of motion
in the discrete Liouville model \cite{tirkkonen}.

\subsection{$\hbar^{-1} = \Gamma_{cusp}(\alpha)$ ?}
Let us comment on the identification of the
Planck constant for the quantization of the KS gravity
as the inverse cusp anomalous dimension inspired
by the BDS anzatz. At the first glance it looks completely groundless
however  the argument supporting this
identification goes as follows. The emergence of the cusp anomaly
in the exponent means that it plays the role of the effective
string tension or equivalently the inverse Planck constant.
Such effective tension emerges if one considers the string whose boundary
is extended along the light-like contours.
It was shown \cite{alday3} that in the limit suggested in \cite{bgk2}
the string worldsheet action can be identified with
$O(6)$ sigma model  and the energy of the ground state in $O(6)$ model
is proportional to the length of the string multiplied
by the $\Gamma_{cusp}(\alpha)$. That is indeed $\Gamma_{cusp}(\alpha)$
plays the role of the effective tension of the string in this special
kinematics. Since in our
case  the boundary of the string worldsheet lies on the Wilson polygon
the effective tension involving the
cusp anomalous dimension is natural.

However certainly this point is far
from being clarified. For instance in the Ward identity for the
special conformal transformation  $\Gamma_{cusp}$ enters as
the multiplier in the anomalous contribution. This claim has been
explicitly checked at the first loops in the gauge theory calculations
and the arguments that it holds true at all orders have been presented.
The  anomalous Ward identity reads as \cite{drum08}

\beq K^{\nu}W(x_1,\dots x_N)= \sum_{i=1}^{n}(2x_i^{\nu}x_i \partial_i -x_i^2 \partial_i^{\nu}) W(x_1,\dots x_N)          =
\frac{1}{2}\Gamma_{cusp}(\alpha)\sum_{i=1}^n ln \frac{x_{i,i+2}^2}{x_{i-1,i+1}^2} x_{i,i+1}^{\nu}
\eeq
and it has been proved
at strong coupling as well \cite{zah}.

In this equation  $\Gamma_{cusp}$ plays the role
of the Planck constant not the inverse one. To match both arguments
we could suggest that in the Ward identity we are considering
the S-dual formulation and therefore the D1 string worldsheet action
instead of the F1 one in $O(6)$ sigma model. This would imply that
the Wilson polygon equivalent to the MHV amplitude could be
considered as the boundary of the D1 string as well. Similarly
the $\Gamma_{cusp}$ enters the loop equation for the Wilson
loops with cusps \cite{gross}
\beq
\Delta_{C}<W(C)>= \sum_{cusps} \Gamma_{cusp}(\alpha, \theta_i)<W(C)>
\eeq
where $\Delta_{C}$ is the Laplace operator in the loop space
and the summation goes over all cusps along the contour C. We assume
that there are no self-intersections.
In this case the $\Gamma_{cusp}(\alpha)$
has the natural interpretation as the inverse Planck constant
since the loop equation has the form of the Schreodinger equation.

In more general setup it is highly desirable to realize the meaning
of the relation of such type  in the first quantized language.
Since the cusp anomalous
dimension is just the renormalization factor for the self-crossing of the
worldline it is very interesting to understand if such self-crossing
is involved into the quantization issue. In particular in the Ising model
the effect of the self-crossing is captured  by the topological term
and in the description of the topological string on $C^3$ somewhat
similar $\theta$  term in six dimensions plays the role of the quantization
parameter indeed \cite{foam}. In the gauge theory language such
objects are related to the renormalization of the double-trace operators
couplings.

Note that generically the relation between  the
YM coupling and string coupling involves the $B_{NS}$ field
\beq
\frac{1}{g_{YM}^2}=\frac{\sqrt{Vol_{T^2}^2 +B_{NS}^2}}{g_s}
\eeq
One could try to speculate that the self-crossing could be
sensitive to the $B_{NS}$ field.
Anyway it is clear that precise identification of the relation between the
quantization
parameter in the KS gravity and the YM coupling is one of the necessary steps
in improving the BDS anzatz.

\section{Integrability behind the scattering amplitudes }
\subsection{General remarks}
In this Section  we shall
discuss the hidden integrability behind the scattering amplitudes and
present the arguments that similarly to the integrability pattern
behind effective actions in N=2 SYM theory  (see \cite{gormir} for the review)
two integrable systems are involved. The degrees of freedom of both
integrable systems are related to the coordinates of the regulator branes.
One of these systems which we identify as the Whitham-like  3-KP one plays the
role of RG flows in the regulator brane worldsheet theory or equivalently
the motion of the regulator brane along the "radial" RG-coordinate.
The second integrable system generalizing the Hitchin-like or spin chain models
involves the effective interactions between the regulator branes. We shall
give arguments that this system is based on the Faddeev-Volkov
solution to the Yang-Baxter equation
for the infinite-dimensional representations of the noncompact $SL(2,R)$ group.

Recall  how  two integrable systems are involved into the description of the
low energy effective actions of $N=2$ SYM theories. The first finite dimensional system
is of the Hitchin or spin chain type and its complex Liouville tori
are identified with the Seiberg-Witten curves.
This spectral curve emerges in the gauge theory
upon the summation over the infinite number of instantons \cite{nekrasov}.

Following \cite{duality} one can
canonically define the dual integrable system whose phase space is built on the
integrals of the motion of the first one. In the simplest
case of $SU(2)$ theory
the phase space for the dual system has the symplectic structure \cite{losev}
\beq
\omega= da\wedge da_{D}
\eeq
where the variables $a,a_{D}$ are the standard variables in N=2 SYM framework
\cite{sw}. The prepotential $\cal{F}$  can be identified with the generating function
of the Lagrangian sub-manifold in the dual system with the $a,a_{D}$ phase space
\beq
H(a(u), \frac{\partial \cal{F}}{\partial a})) =u
\eeq
and obeys the Hamilton-Jacobi equation
\beq
 \frac{\partial \cal{F}}{\partial log \Lambda}=H
\eeq
In the brane setup the prepotential defines the semiclassical "wave function"
of the D4 brane $\Psi(a)\propto exp(\hbar^{-1} {\cal{F}}(a))$
in the IIA brane picture where perturbatively the argument
of the wave function can be identified with  coordinate
of the D4 brane on the NS5 brane. The total  perturbative
prepotential in $SU(N_c)$ can be considered as a sum of the exponential
factors in the product of the wave functions of  $N_c$
D4 branes. At the A-model side these wave functions can be considered
in the Kahler gravity framework and the arguments of the wave function
have to be treated as the Kahler classes of the blow-upped spheres.

The integrals of motion provide the moduli space of the complex structures
in the Calabi-Yau geometry in the B model hence
we are precisely in the KS framework. In this B-model
formulation we consider the argument of the brane wave function
as the coordinate on the moduli space of the complex structures.
The dual Whitham-type integrable system naturally
defines the $\tau$-function of the 2d Toda theory
formulated in terms of the
chiral fermions on the Riemann surface with two marked points.
Upon  perturbing $N=2$ theory down to $N=1$ the moduli space disappears
and the number of vacua becomes finite. In the integrability framework
this is treated in the following manner. The Hamiltonian of the first
finite-dimensional system turns out to coincide with the
superpotential of the $N=1$ system \cite{hollowood}. That is the vacua of the
gauge theory at the classical level correspond to the equilibrium
points in the Hamiltonian system $W'=0$.

\subsection{3-KP system}

Let us turn to the integrable structure relevant for the scattering
amplitudes at generic kinematics and
first  identify the degrees of freedom
and evolution "times". As we have described
above   the fermionic degrees of freedom correspond to the
noncompact branes  localized on the Riemann surface. The
two-dimensional field theory corresponds to the reduction of the Kodaira-Spencer
theory on the two-dimensional surface. The coordinate on the Riemann surface
is related with the coordinate on the moduli space $M_{0,4}$. The Kodaira-
Spencer theory is described by the two dimensional Lagrangian
\beq
L_{KS}= \int (\partial \phi \bar{\partial} \phi +\frac{1}{\lambda}
\omega \bar{\partial}\phi + \frac{\lambda}{\omega}(\partial \phi)^2 \bar{\partial}\phi)
\eeq
where $\phi$ is the basic scalar in KS theory $\omega$ is one-form on the surface
and $\lambda$ is the topological string coupling constant.
It was argued recently  \cite{dv07} that the cubic interaction term  in the KS
Lagrangian can be formulated as the screening operator in the two-dimensional conformal theory.
The fields on the surface are in the
external abelian connection of the Berry type which tells how the B - branes
transform under the change of the complex structure fixed by the momenta
of external particles.

As we have mentioned in the
$c=1$ example there are two possible set of "times", "compact" and "noncompact" ones.
The compact ones correspond to the variation of the complex structure at infinities
and are responsible for  the insertion of the vertex operators of the "tachyonic" degrees
of freedom while the noncompact ones  correspond to the insertions of the noncompact
B-branes at the particular values of the cross-ratios. The gluon vertex operators in this
framework correspond to the tachyonic vertex operators in $c=1$ model.
The set of Kontsevich  times determined by the positions
of the B-branes  are defined by (12)
where $z_{i}$ are the corresponding cross ratios.

The form of the Riemann surface $H(u,v)=0$
dictates that there are three infinities and therefore we are dealing with the
particular solution to 3-KP  integrable system.
To describe the integrable system it is convenient
to introduce the chiral fermions
with the following mode expansion
\beq
\psi(x_i)= \sum_{n}\psi^{i}_{n+1/2}x_i^{-n-1},\qquad
\psi^{*}(x_i)= \sum_{n}\psi^{*i}_{n+1/2}x_i^{-n-1}\eeq
around the i-th infinity, $i=1,2,3$, and the
commutation relations
\beq
\{\psi_{n}^{i},\psi_{m}^{*j}\}=\delta^{ij} \delta_{n+m,o}
\eeq
Defining the vacuum state by relations
\beq
\psi_n|0>=0, \quad \psi_n^{*}|0>=0, \quad n>0
\eeq
 the generic state $|V>$ can be presented in the form
 \beq
 |V>=exp(\sum_{i,j} \sum_{n,m} a_{nm}^{ij}\psi^{i}_{-n-1/2}\psi^{*i}_{-m-1/2})|0>
 \eeq
 where the point of Grassmanian representing the topological vertex
 was derived in \cite{foam}. for instance the diagonal
 coefficients read as
 \beq
 a^{ii}_{nm}=(-1)^n \frac{q^{m(m+1)-n(n+1)}}{[m+n+1][m][n]}
 \eeq

The tau-function of the 3-KP system plays the role of the generating function
for the  MHV amplitudes.
In the semiclassical approximation we can safely consider
the differential on the classical Riemann surface
\beq
dS=vdu
\eeq
which yields the semiclassical brane wave function
\beq
\Psi_{qs}\propto exp(- \hbar^{-1} \int ^{x} v(u)du)
\eeq
involving the dilogaritms. The tau-function obeys the
3-KP equation and there are the additional
$W_{1+\infty}$ Ward identity written in terms of the fermions
\beq
\oint_{u}\psi{*}(u)e^{nu}\psi(u) +
(-1)^n \oint_{v}\psi{*}(v)e^{nv}\psi(v) +
\oint_{s}\psi{*}(s)e^{ns}\psi(s)=0
\eeq
where the sum over three asymptotic regions is considered.

The quantization of the system can be done most effectively
in terms of the Baxter equation.
The  Baxter equation implies that the regulator branes
are localized on the surface. Hence the whole set of the
equations determining amplitudes involves the dual
conformal transformations on the regulator worldvolume
and the set of Ward identities  for the coordinate
of regulator brane in the transverse moduli space. It is these
Ward identities which fix the dependence of the amplitude
on the conformal invariants for large number of external legs.

The precise form of higher Hamiltonians from $W_{1+\infty}$ responsible for the
higher conservation laws in the scattering amplitude problem can
be written as the fermionic bilinears \cite{vafa}.
Generically as was discussed in \cite{vafa} one
has some unbroken part of $W_{\infty}$ which
annulates the $\tau$-function corresponding to the
topological vertex and therefore
the scattering amplitude in the form of BDS-like anzatz.

\subsection{On the Faddeev-Volkov model}
Let us turn now to the description of the second integrable system
representing the particular solitonic sector of the infinite-dimensional
integrable system.
We shall conjecture that the integrable system at the generic kinematics
is the generalization of the $SL(2,C)$ spin chain relevant for the
Regge limit of the amplitudes.

The finite-dimensional integrable systems can be usually defined in terms of the R-matrix.
The Faddeev-Volkov model is defined via the Drinfeld
solution  to the Yang-Baxter equation which provides
the  universal R-matrix acting on $U_{q}(SL(2,R))\otimes U_{\tilde{q}}(SL(2,R))$.
The corresponding statistical model describes the discrete quantum Liouville theory \cite{bms} with
the following partition function
\beq
Z=\int \prod_{ij}W_{p-q}(S_i-S_j)\prod_{kl}\bar{W}_{p-q}(S_k-S_l)\prod_{i} dS_i
\eeq
where the Boltzmann weights depend only on the differences of the spins $S_k$
at the neighbor cites and rapidity variables
at the ends of the edge. The first product is over the horizontal
edges $(i,j)$ while the second product is over the vertical edges $(k,l)$. The integral
is over all internal spin degrees of freedom. In the fundamental R-matrix
the cross-ratios of the relative rapidities of the particles play the
role of the local inhomogeneities in the lattice model and Boltzmann weights
are defined as \cite{bms}
\beq
W_{\theta}(s)=F(\theta)^{-1}e^{2\eta \theta s}\frac{\Psi(s+ic_b \theta \pi)}
{\Psi(s-ic_b \theta \pi)}
\eeq
where spin and local rapidity variables are combined together in  the argument of the function
$\Psi$
and $F(\theta)$ is some normalization factor. The relative importance
of the spin variables and the local inhomogeneities depends on the
value of the YM coupling constant and the kinematical region.

Semiclassically when $b \rightarrow 0$ the spin variables are frozen  and the Boltzmann weight behaves as
\beq
W_{\theta}(\rho/2\pi b))=exp(-\frac{A(\theta|\rho)}{2\pi b^2} + ...)
\eeq
where
\beq
A(\theta|\rho)=iLi_2(-e^{\rho-i\theta}) -i Li_2(-e^{\rho+i\theta})
\eeq
The extremization of the semiclassical action yields
the Bethe Anzatz type equations connecting the dynamical spin variables
with the local rapidities
\beq
\prod_{i} \frac{e^{\rho_i}+e^{\rho_j +\theta_{ij}}}{e^{\rho_j}+e^{\rho_i +\theta_{ij}}}=1
\eeq

Let us try to compare the brane geometry behind two integrable systems
behind the low-energy $N=2$ SYM and in the $N=4$ scattering problem. In $N=2$ case
in the IIA picture we have $N_c$ D4 branes stretched between
two NS5 branes and coordinates of D4 branes on the NS5 brane
correspond to the vacuum expectation values of the scalars. The
second set of degrees of freedom is provided by the
low-dimensional branes on D4 branes with attached strings connecting D4 branes.
The first "fast" integrable system describes the dynamics of the
strings while the  second "slow" integrable system describes
the dynamics of D4 branes on the moduli space of the vacua. Very
similarly in the scattering geometry the D4 branes are substituted
by the B-model branes localized on the moduli space $M_{0,4}$
and their dynamics is described by the Whitham-type 3-KP hierarchy while the
second integrable system with N degrees of freedom corresponds
to the dynamics of the open strings  attached
to the B- branes.

Completing this Section let us make  comment on how the
interplay between "soft" and "regulator"
degrees of freedom is captured by the integrable dynamics. To this aim
remind the description of the KdV hierarchy in terms of the
Liouville field. The KdV equation can be considered
as the rotator on the coadjoint orbit of the Virasoro algebra.
The coadjoint orbit is the symplectic manifold  and  the geometrical action on
the Virasoro orbit is the Liouville action \cite{shatashvili}. It
can be derived upon integration of the chiral fermion in the
external gravitational field. On the other hand the KdV Hamiltonians
are provided by the integration of the "heavy" non-relativistic
degree of freedom in the same gravitational field that is $\log
det(d^2 +T)$, where $T$ - is the two-dimensional
energy stress tensor. In the Lax representation we consider the isospectral
evolution of the Baker-Akhiezer function which is the eigenfunction
of the Schreodinger operator and can be attributed to the "heavy" degree
of freedom. This is similar to our case since the wave functions of the
noncompact branes can be considered as the BA functions of the integrable
system.

\section{Comments on the Regge limit}
In this Section we shall discuss some features which hopefully
could help in the explanation
of  the Reggeization of the amplitude. The
interpretation of the Reggeon in the dual picture was discussed
in \cite{gkk} where its identification as the singleton
representation in $AdS_3$ was suggested and the universality class
of the multireggeon system was clarified. The dual picture
behind the pomeron state was discussed in \cite{strassler}.
We shall conjecture on the interpretation  of the reggeon
degrees of freedom in the KS gravity framework.

Remind that the Reggeon field $V(x)$ enters the effective
Lagrangian being coupled to the semi-infinite light-like
Wilson line \cite{lipatoveff} playing the role of the source
\beq
L_{int}=-\frac{1}{g}\partial_{+} P exp(-\frac{g}{2}\int_{-\infty}^{x^{+}} A_{+}dx_{-})
\partial^2 V_{-}
-\frac{1}{g}\partial_{-} P exp(-\frac{g}{2}\int_{-\infty}^{x^{-}} A_{-}dx_{+})
\partial^2 V_{+}
\eeq
where $x_{+}, x_{-}$ are the light-cone coordinates and $A$ is the
conventional gluon field.
That is according to the gauge/string
duality it is natural  to lift the reggeon field to the
field in the bulk. Hence  the correlator of the light-like Wilson lines
could be derived by differentiation of the bulk reggeon action with
respect to the  boundary values of the reggeon field. It is this line of
reasoning was implied in
\cite{gkk} when interpreting the Reggeon as the singleton in the bulk action.

The reggeon field does not transform under the local gauge
transformation however carries the global color charge
and therefore interacts with the conventional gluon.
It is this interaction amounts to the BFKL hamiltonian
governing the $t=log s $ evolution of the pomeron state
\cite{bfkl} and the corresponding miltireggeon BKP
generalization \cite{bkp}. Note that the situation
is reminiscent to the standard interplay between
the local and global symmetries in the brane picture.
In the color $N_c$ branes worldvolume theory the gauge symmetry
on the
flavor $N_F$ branes is seen as the global flavor
symmetry. The open strings connecting the color and flavor
branes carry the global flavor number.

We could conjecture that the reggeized gluon can be
identified with the open string stretched between
two regulator branes that is it can be considered
as the massive vector "gauge" particle for the "flavor"
gauge group on the set of the IR regulator  branes.
Such reggeon field indeed plays the role of the source on the
regulator brane worldvolume and therefore the Wilson polygon
on the regulator brane worldvolume presumably can be derived upon
the differentiating the classical action in the bulk over
the boundary values of the reggeon field.

The reggeized gluon emerges upon the re-summation
of the perturbative series hence one could try to identify the
particular limit in the KS framework which could provide
the multireggeon dynamics of the all-loop amplitude in the generic kinematics.
To this aim it is useful to compare  the integrable structures at the generic
kinematics we discussed above and the one responsible for the
Regge limit \cite{lipatov1,fadkor}.

The Regge limit is described in terms of the
$SL(2,C)$ spin chains when the number of sites
in the chain corresponds to the
number of reggeons. The possible limit which could
yield such spin chain from the Faddeev-Volkov model
or statistical model \cite{bms}
looks as follows. In the model \cite{bms} the statistical weights
depend on the sum of the local rapidities  and the spin variables.
It is clear that one can not expect the semiclassical limit
of the quantum dilogarithm to be relevant since the reggeization
of the gluon happens upon   the nontrivial resummation of
the perturbation series.

Fortunately there is the limit \cite{bms} corresponding to the strong
coupling region in the Liouville theory when
the quantum dilogarithms  reduce to the ratio of gamma
functions depending on the $SL(2,R)$ spin variables
\beq
\Psi_{c_b\rightarrow 0}(s+\eta x)\propto \frac{\Gamma(1-s+ix/2)}{\Gamma(1-s-ix/2)}
\eeq
where $|b|=1$ and $x $ is the rescaled local rapidity.
The leading argument depends on the difference
of two infinite-dimensional representations in the neighbor
sites and the expression coincides with the
fundamental R-matrix involved into the
$SL(2,R)$ spin chains.  That is in this particular limit we
get the statistical weights
or R-matrixes depending only on the $SL(2,R)$ spins similar
to the BFKL-type Hamiltonian while  the local rapidity  yields
the "time" variable $log s$. Note that clearly this
suggestive argument needs for further clarification.

Another possible limit which can be compared with
is the semiclassical limit of the multi-reggeon
system which is described
in terms of the higher genus Riemann surface of the type
\beq
y^2=P_{N}^2(x)- 4x^{2N}
\eeq
where N- is the number of Reggeons and $P_N$ is the
N-th order polynomial depending on the higher integrals of motion.
It  was shown
\cite{gkk} that the Reggeon system belongs to the same universality class
as N=2 SYM with $N_f=2N_c$ at strong coupling orbifold point.
In that case the brane geometry behind the low-energy
effective action is known \cite{witten} and the theory is
realized on the M5 brane with worldvolume $(R_{3,1}, \Sigma)$
where the surface $\Sigma$ lies in the internal space.

In the scattering geometry it is known \cite{gkk} that the spectral curve
of the integrable spin chain is embedded into the
complexified $(x,p)$ space where $x$ is the coordinate in the conventional
Minkowski space. On the other hand we have discussed the geometry
in the internal space $(u,v)$ which can be roughly
thought of as the $T^{*}M_{(0,4)}$. The two viewpoints can be matched
if we use the realization of the $T^{*}M_{(0,4)}$ as the $T^{*}(M^c//T)$
hence we indeed can try to treat the spectral surface as the sub-manifold
in the complexified phase space.

Some comments on the role of $SL(2,C)$ group is in order.
It is just the group of the Lorentz rotations which act
both in the coordinate and the momentum space. In the A model
the set of Lagrangian branes yields the knots and there
is the natural action of $SL(2,C)$ on the knot complement.
That is the $SL(2,C)$ holonomies around the boundary torus
yields the degrees of freedom in the spin chain model.
In the B model side one can try to relate the
group with the $SL(2,R)$ structure which has been considered
within the lifting of  KS theory on the Riemann surface
to the three-dimensional CS theory. The derivation
of the spin chain system from the set of Wilson lines
in the CS theory has been discussed long time ago
\cite{witten89} and the similar derivation
could be expected here as well.

\section{Discussion}

In this paper we have suggested the relation  between the multiloop
MHV amplitudes and  effective gravity on the moduli spaces
provided by the kinematical invariants of the scattering particles.
This viewpoint  allowed us to suggest the relevant integrability pattern
and amplitudes were treated as the fermionic current correlators on the
moduli spaces.
The key idea is that the scattering process induces
the back-reaction on the geometry of the "momentum space" through
the nontrivial dynamics on the emerging moduli space.
That is one can  say that the tree
amplitude is dressed by the effective gravitational degrees of freedom
which can be
treated within the Kahler gravity  in the A type geometry or
KS gravity in the type B model. They  are
identified with the coordinates of Lagrangian branes in the A model or the
corresponding noncompact
branes in the B model. These branes serve as the effective IR regulators
in the theory. On the field theory side the
four fermion currents   correlator on the moduli space  is identified with the two-mass
easy box amplitude which is the basic block in the whole answer.
Within the conventional calculation of the Feynman diagrams the relevant moduli spaces
are parameterized by the Schwinger or Feynman parameters.

The BDS anzatz corresponds to the
semiclassical limit in the effective gravity and $\Gamma_{cusp}$ has
to be identified with the effective  inverse Planck constant in KS gravity.
The anzatz has to be modified and our proposal suggests
several natural directions of its generalizations. First, one could imagine that
the quantization parameter can be generalized to more complicated
function than the cusp anomalous dimension
which would respect the S-duality of $N=4$ theory. The next
evident point concerns the  full quantum
theory in the gravity framework which effectively substitutes the
dilogarithm function in the BDS anzatz by the quantum dilogarithm.
However these modifications do not produce proper higher polylogarithms
which are known to appear in higher loop calculations of the
amplitudes and Wilson polygons. The most natural way to get desired
higher polylogarithms in our picture is to take into account the
nontrivial Feynman diagrams in the two-dimensional KS theory
probably involving loops. Indeed increasing the number
of vertexes in the KS tree diagrams we increase the
transcendentality of the answer. We expect that all mentioned
generalizations are necessary to be taken into account
to get the correct all-loop
answer.

We have identified the most natural integrable structure
behind the scattering amplitudes which are considered as
a kind of the "wave functions" in the particular model.
The KS gravity in our case naturally involves
the 3-KP hierarchy and the roles of the "time" variables
are played by the combination of the conformal cross-ratios
expressed in terms of the external momenta.
The second finite-dimensional integrable system is conjectured
to be related to the Faddeev-Volkov model however this point
deserves for further investigation.
The integrability is responsible for the
conservation laws in addition to the dual superconformal
symmetry.
The relevant  Ward identities
correspond to the area preserving symplectomorphysms of the spectral
curve similar the considerations discussed previously in $c=1$ model.

The additional IR regulator branes
added into the picture
are responsible for the blow up of the internal
momentum space in the manner dictated by
the scattering process. The blow up of the internal geometry
physically corresponds to the IR regularization of the field theory and
the anomaly in the transformations
in the momentum space tells that the
regularization does not decouple completely. This a
little bit surprising picture implies that  we  have to take into
account the dynamics of the regulator degrees of freedom as well.
It is highly desirable to develop the microscopic derivation
of these IR branes. One can imagine that such branes emerge upon
the peculiar summation of the noncommutative instantons
in the effective abelian target space description of
our simplified $C^3$ geometry.

Naively IR regulators are treated semiclassically but generically the
fermion currents representing the
regulator branes obey the quantum  Baxter equation. It is
clear that the discrete Liouville model plays the important
role in the whole picture providing the particular
gravitational dressing of the operators involved. We expect
that these discrete Liouville modes correspond to the
remnant of the reparametrization of the Wilson polygons coming from the cusps.
The regulator branes to some extend play the role  similar to the Liouville
walls in the $c=1$ model.

We expect that our treatment of the scalar box function imply
that the hidden structure behind the gauge box diagrams holds for
the non-MHV case as well.
In particular we expect that non-vanishing all-plus
amplitude in the usual YM theory which anticipated to be of the
anomalous nature since long corresponds  to the purely anomalous
part of the algebra of the symplectomorphysms of the
spectral curve.

One of the most inspiring findings of the paper is the
appearance of the hidden  "new massive degree of freedom". They correspond
on the A model side to the D2 brane wrapped around the 2-cycle created by the
scattering states or the open string stretched between
two IR regulator branes in the B model. It is somewhat similar
to the W-boson or monopole states in the Seiberg-Witten description of
low-energy effective action of $N=2$ theory
however  its "mass" is fixed by the kinematical invariants
of the scattering particles. It would be very interesting
to develop these reasoning further and determine
the corresponding walls of marginal stability in the space of the
kinematical invariants.
In the Regge limit  we anticipate its important role in the Reggeon field theory.
We plan to elaborate this issue further elsewhere.

It is evident that our proposal requires
clarifications in many respects. In particular the clear understanding
of the amplitudes of the gluon scattering with  generic chiralities
is absent yet and our conjecture for the improvement
of the BDS anzatz deserves for  further evidences. Nevertheless we
believe that the dual representation of amplitudes in terms of the
dynamical systems on the moduli space of the regulator branes
we have  suggested is the useful
step towards the derivation of the dual geometry
responsible for the summation of the perturbative series in SYM theory.

\section*{Acknowledgements}
I would like to thank J. Drummond, E. Gorsky, D.
Gross, S. Gukov, G. Korchemsky, A. Levin, V. Mikhailov,
N. Nekrasov, A. Rosly,  M. Shifman,
E. Sokatchev, A. Tseytlin , A. Vainshtein, M. Voloshin , V. Zakharov
and A. Zhiboedov for the
useful discussions. The parts of the
work have been done during the programs " Strong Fields, Integrability and
Strings" at
INI, Cambridge, "From strings to things" at INT, Seeatle  and
"Non-Perturbative Methods in Strongly Coupled Gauge theories" at
GGI, Florence. I would like to thank the organizers for
hospitality and support. I would like to thank also IHES, SUBATECH,
FTPI at University of Minnesota and Universite Paris XI
for the kind support while the parts
of the work were done.
The work  was supported in part by grants
RFBR 09-02-00308 and PICS- 07-0292165.


\section*{Appendix}

To discuss the multi-loop calculations it is  useful to utilize the geometrical
picture behind the one-loop calculations which we shall
review  following \cite{dav}.
There exists the explicit map of the box diagram to the
hyperbolic volume of the particular simplex build from the kinematical
invariants of the external momenta.  Introduce
the Feynman parametrization of the internal generically massive propagators
with the parameters $\alpha_i$.  If one considers the one-loop N-point function
with the external momenta $p_i$
in D space-time dimensions it can be brought into the usual form
\beq
J(D,p_1,\dots p_N)\propto
\int_{0}^{1} \dots \int_{0}^{1}\prod d\alpha_i \delta(\sum \alpha_i -1)
[\sum \alpha_i^2 m_i^2 +\sum_{j<l} \alpha_i \alpha_j m_j m_l C_{il}]^{D/2-N}
\eeq
where
\beq
C_{jl}=\frac{m_i^2 +m_l^2 -k_{jl}^2}{2m_j m_i},\qquad k_{ij}= p_i -p_j
\eeq
and $m_i$ is the mass in the i-th propagator.

It is possible \cite{dav}  to organize for the generic
one-loop diagram the $N$ dimensional simplex defined as follows.
First introduce  the $N$ mass vectors $m_i a_i$ , where $a_i$ are the unit vectors. The
length of the side connecting the i-th and j-th mass vectors is $\sqrt {k_{ij}}$
that is one can define the momentum side of the simplex. Therefore the $N$-dimensional simplex involves
$\frac{N(N+1)}{2}$ sides including
N mass sides as well as $\frac{N(N-1)}{2}$ momentum sides. At each vertex $N$ sides meet and at all vertices
but one there are one mass side and $(N-1)$ momentum sides. The volume of such $N$-dimensional simplex
is given as follows
\beq
V^{(N)}=\frac{(\prod m_i)\sqrt {det C}}{N!}
\eeq
There are $(N+1)$ hypersurfaces of dimension $(N-1)$ one of which contains only momentum sides
and can be related with the massless N-point function.

Upon  the change of variables the loop integral get transformed
into the following form
\beq
J (D,p_1,\dots p_N)\propto \prod m_i^{-1}
\int_{0}^{\infty} \dots \int_{0}^{\infty}\prod d\alpha_i \delta(\alpha^{T} C \alpha -1)
(\sum \frac{\alpha_i}{m_i})^{N-D}
\eeq
It is useful to introduce the content of the $N$-dimensional solid angle $\Omega^{(N)}$
subtended by the hypersurfaces at the mass meeting point. It turns out that $\Omega^{(N)}$
coincides with the content of the $(N-1)$ dimensional simplex in the hyperbolic space
whose sides are equal to the hyperbolic angles $\tau_{ij}$ defined at small masses as follows
\beq
C_{ij}=cosh\tau_{ij}
\eeq
Then the integral for the case $D=N$ acquires the following form
\beq
J(N,p_1,\dots p_N)=i^{1-2N}\frac{\pi^{N/2}\Gamma(N/2) \Omega^{(N)}}{N! V^{(N)}}
\eeq
hence the calculation of the Feynman integral is nothing but the
calculation of the hyperbolic volume in the proper space. The case
$N\neq D$ can be treated similarly with some modification \cite{dav}.

To avoid  IR divergence  it is useful to start with
the box diagram with all off-shell particles that is
$D=N$ simplices in the hyperbolic space.
\beq
J(4,p_1,p_2,p_3,p_4)= \frac{2i\pi ^2 \Omega^{(4)}}{m_1 m_2 m_3 m_4 \sqrt{det C}}
\eeq
Since  all internal propagators are massless in our case
we get the ideal hyperbolic tetrahedron whose all vertices are
at infinity. In the massless limit we get
\beq
(m_i^2 m_2^2 m_3^2 m_4^2 det C)_{m_i\rightarrow 0}=
\frac{1}{16} \lambda(k_{12}^2 k_{34}^2, k_{13}^2 k_{24}^2, k_{14}^2 k_{23}^2)
\eeq
where the K\"{a}llen function $\lambda(x,y,z)$ is defined as
\beq
\lambda(x,y,z)=x^2 +y^2 +z^2 -2xy -2yz -2zx
\eeq
and $\sqrt{-\lambda}$  is just the area of the triangle with sides
$\sqrt{k_{12}^2k_{34}^2}, \sqrt{k_{23}^2k_{24}^2}, \sqrt{k_{31}^2k_{23}^2}$. The hyperbolic volume
of the ideal tetrahedron under consideration reads as
\beq
2i\Omega^{(4)}= Cl_2(\psi_{12}) +Cl_2(\psi_{13}) +Cl_{2}(\psi_{23})
\eeq
where the dihedral angles are defined via the kinematical invariants
\beq
-cos \psi_{12}= \frac{k_{13}^2 k_{24}^2 + k_{14}^2 k_{23}^2 - k_{12}^2 k_{34}^2 }
{\sqrt{k_{13}^2 k_{23}^2 k_{14}^2 k_{43}^2 }}
\eeq
\beq
-cos \psi_{13}= \frac{k_{14}^2 k_{23}^2 + k_{12}^2 k_{43}^2 - k_{13}^2 k_{24}^2 }
{\sqrt{k_{14}^2 k_{23}^2 k_{12}^2 k_{43}^2 }}
\eeq
\beq
-cos \psi_{14}= \frac{k_{12}^2 k_{34}^2 + k_{13}^2 k_{24}^2 - k_{14}^2 k_{32}^2 }
{\sqrt{k_{13}^2 k_{24}^2 k_{12}^2 k_{43}^2 }}
\eeq
and $\psi_{12}=\psi_{34},\quad \psi_{13}=\psi_{24},\quad \psi_{14}=\psi_{32}$.
The functions involved are defined as
\beq
Cl_2(x)=Im [Li_2(e^{ix})= - \int_{0}^{x} dy ln|2sin y/2|
\eeq
In the case of the two mass-easy box diagram defining the one-loop
MHV amplitude the additional simplification of the kinematical invariants
happens since two external particles are on the mass shell. In this case
the arguments of the $Li_2$ function degenerate to the conformal
ratios of four points.

\end{document}